# Cloud Infrastructure Service Management – A Review

A. Anasuya Threse Innocent

Department of Computer Science and Engineering, SCT Institute of Technology, Visvesvaraya Technological University
Bangalore, Karnataka, India
*anasuya.smith@gmail.com*

**Abstract**

The new era of computing called Cloud Computing allows the user to access the cloud services dynamically over the Internet wherever and whenever needed. Cloud consists of data and resources; and the cloud services include the delivery of software, infrastructure, applications, and storage over the Internet based on user demand through Internet. In short, cloud computing is a business and economic model allowing the users to utilize high-end computing and storage virtually with minimal infrastructure on their end. Cloud has three service models namely, Cloud Software-as-a-Service (SaaS), Cloud Platform-as-a-Service (PaaS), and Cloud Infrastructure-as-a-Service (IaaS). This paper talks in depth of cloud infrastructure service management.

## 1. Introduction and Cloud Basics

1.1 Introduction

According to NIST [1], Cloud computing is a model for enabling ubiquitous, convenient, on-demand network access to shared pool of configurable computing resources that can be rapidly provisioned and released with minimal management effort or service provider interaction. Cloud model promotes five essential characteristic, three service models, and four deployment models. The five essential characteristics are; on-demand self-service, broad network access, resource pooling, rapid elasticity, and measured service. The three service models are; Cloud Software-as-a-Service (SaaS), Cloud Platform-as-a-Service (PaaS), and Cloud Infrastructure-as-a-Service (IaaS), together called as the SPI model [1, 2].

1.2 Infrastructure-as-a-Service

IaaS refers to computing resources as a service. IaaS provides virtual machines, virtual storage, networking technology, data center space, and other hardware assets as resources that clients can provision. The IaaS service provider manages the entire infrastructure, while the client is responsible for the deployment aspects. Examples of IaaS service providers are; Amazon elastic Compute Cloud (EC2), Clever, Eucalyptus, GoGrid, FlexiScale, Linode, Nimbus, Open Nebula, PerfCloud, RackSpace Cloud, and Terremark [3, 4]. Companies which carry out research-intensive projects can us IaaS in a cost effective manner, as the new computing infrastructures required for analysis and testing need not be installed in the company premises. Storage-as-a-Service (StaaS) or the data Storage-as-a-Service (dSaaS) is one of the sub-services provided by the IaaS. EMC, Sun, IBM, Microsoft, Amazon, and Google are the major providers of Storage-as-a-Service.

1.3 Platform-as-a-Service

The Platform-as-a-Service sits on top of Infrastructure-as-a-Service in the SPI model. PaaS service adds integration features, middleware, and other deployment capability services to the IaaS. Examples of PaaS services are; Force.com, GoGridCloudCenter, Google AppEngine, AppJet, Etelos, Qrimp, and Windows Azure Platform [3, 5]. The new variety of PaaS, Open Platform-as-a-Service provides no constraint on choice of development software and avoids the possibility of lock-in.

1.4 Software-as-a-Service

SaaS offers built-in business applications which allow users to run applications remotely from the cloud on a pay-as-you-go model. Examples of SaaS cloud service providers are; GoogleApps, Oracle On Demand, SalesForce.com, SQL Azure etc. [3] SaaS is based on Application Service Providers (ASPs), and as an enhancement of ASP model two modes of SaaS namely *Simple Multi-tenancy* and *Fine Grain Multi-tenancy* are available. In simple multi-tenancy, each customer has their own resources segregated from other users. In fine grain multi-tenancy, all resources are shared except customer data and access capabilities. The massively scaled Software-as-a-Service keeps the cost per user very less by allowing millions of users to use the same service at the same time. Environments such as Facebook, eBay,

Skype, GoogleApps etc., are designed for massive scaling [5]. In a nutshell, SaaS is a complete operating environment with applications, management, and the user interface.

1.5 Cloud Deployment Models

Peter et al. through the NIST definition of Cloud computing [1] displays the four deployment models of cloud computing. If the cloud infrastructure is operated solely for an organization, managed by the organization or by a third party, it is called as the *Private cloud*. The Private cloud or the *internal cloud* provides full control over data, security, and QoS. If the cloud infrastructure is shared by several organizations and if it supports a specific community having shared concerns, and managed by the organizations or by a third party, then it is called as a *Community cloud*. *Public cloud* or the *external cloud* is made available to the general public on a pay-as-you-go fashion, and is owned by an organization selling cloud services. In the public cloud, computing resources are dynamically provisioned over the Internet via Web applications or Web services from an off-site third party provider. Hybrid cloud infrastructure is a combination of two or more clouds (private, community, or public) that remain as unique entities, but are bound together by standardized technology which enables data and application portability between the clouds [1, 2].

This paper is structured in a manner that, Section 2 describes about the cloud infrastructure models; Section 3 depicts the various cloud infrastructure management models; Section 4 pictures the new integrated cloud service models and services.

## 2. Cloud Infrastructure

Autonomic systems, with the high level guidance from humans autonomically decide what steps to be done to keep the system stable; and constantly adapt themselves to changing environmental conditions. Just like the biological systems, autonomic systems maintain their state and adjust operations considering changing components, workload, external conditions, hardware, and software failures. MAPE (Monitoring, Analysis, Planning, Execution) is a well-known autonomic control loop. To adapt to cloud specific solutions an extended MAPE-K loop, called A-MAPE-K was introduced by Michael et al. [6] A in A-MAPE-K stands for the *adaptation phase* and K for the *knowledge management phase* defined. The adaptation phase is used to achieve balance to the virtualization layer. Case Based Reasoning is a feasible knowledge management technique to be used for autonomic management of cloud infrastructure.

The IaaS cloud computing is always an attractive computing concept in enterprises from viewpoints of both a *server user* who makes use of servers and a *server administrator* who prepares servers for server users and manages them. IaaS cloud enables on-demand provisioning of computational resources in the form of Virtual Machines (VMs) deployed in a cloud provider's data center minimizing or even eliminating associated capital costs for cloud consumers and allows the server users to add or remove capacity from their IT infrastructure to meet peak or fluctuating service demands while paying only for the actual capacity used. The working of cloud infrastructure services can be summarized as follows: (i) Server administrator creates virtual server templates and enrolls them as service catalogs into a cloud manager in advance using a cloud manager portal; (ii) Server users open a cloud service portal and select an appropriate service catalog when they wish to use the servers; (iii) Service catalog request sent from the service portal to the cloud manager; (iv) Cloud manager creates a new virtual server on cloud resource pool by copying the template linked with the requested catalog and informs the users that the requested virtual server is ready; (v) Server administrator manages the virtual servers on the cloud management portal and the server users can directly use the virtual servers [2].

2.1 IaaS Providers

The highest profile IaaS provider is *Amazon's Elastic Compute Cloud, EC2*. It supports Linux,Sun Muicrosystems' OpenSolaris and Solaris Express Community Edition, Microsoft's Windows Server 2003, and most of the common operating systems except IBM and HP (they have their own clouds). The *Elastic Block Storage (EBS)* provides persistent storage in EC2. Also, *Simple Storage Service (S3)* of Amazon can also be used independently for the storage purpose. EC2 charges in two types; hourly charge per virtual machine, and data transfer charge.

*Rackspace Cloud* provides the services much closer to Internet Service Provider (ISP) than EC2. It charges for storage space, bandwidth usage, and compute cycles.A GoGrid provider emphasizes ease of use and offers technical control including load balancing than either EC2 or Rackspace cloud. GoGrid charges for; storage space more than 10 GB, outbound data transfer, server RAM hours [5].

*Eucalyptus* – Elastic Utility Computing Architecture for Linking Your Programs To Useful Systems is a system for implementing on-premise private and hybrid clouds using the hardware and software infrastructure that is in place without modification. Eucalyptus platform have three level hierarchical architecture; top level is the Cloud Controller (CLC) node responsible for decision making, middle one is Cluster Controller (CC) responsible for keeping track of resource usage in its cluster, and the bottom one is the Node Controller (NC) having two main responsibilities; monitoring resource usage, and managing virtual resources [7].

2.2 Inside Cloud Infrastructure

Cloud infrastructures employ a *virtualization layer* to ensure resource isolation and abstraction. They are designed to provide restricted visibility to both users and IaaS providers. The Virtual Machine (VM) instances are blocked from looking down into the infrastructure and the IaaS providers are not allowed to look inside the running VM instance. In cloud environments, it is not clear that who is responsible for fixing problems. Some of the cloud providers run online support forum, example - Amazon EC2 Online forum. Most of VM problems in cloud infrastructure are due to configuration events. The *CloudInsight* is a solution proposed by Ahsan et al. [8] monitors and tracks the configuration attributes of VM instances and infrastructure, and uses the historical event records to determine the root cause of problematic VM instances, and further provides a check-and-solves troubleshooting process to resolve user reported problems automatically. It is a monitoring mechanism at the infrastructure level, and the event data is structured and stored in a Configuration Management Database (CMDB). CloudInsight aims to solve connectivity, infrastructure, and performance related problems by automating the problem reasoning process and has an interactive troubleshooting procedure.

## 3. Infrastructure Management

Cloud infrastructure is provided to the users in a more scalable and elastic way by IaaS providers through pay-per-use business model in the form of virtual machine. The problem is how to manage and monitor the IT infrastructure in the cloud. The concerns are; vendor lock-in, security, availability, portability etc. To overcome these problems, different infrastructure management models/projects are developed. Some of them are discussed in the following sections.

3.1 RESTful Cloud Management System (CMS)

*REprentational State Transfer (REST)* was introduced by Roy Fielding in the year 2000 [9]. REST, architectural style for distributed hypermedia systems describing the software engineering principles and the interaction constraints. Fundamental principles of REST are: (i) each resource is referenced using a uniform resource identifier (URI), (ii) resources are manipulated by only four major HTTP methods; GET, PUT, DELETE, and POST, (iii) each interaction with a resource is stateless. REST can enhance existing management systems since resources in REST can model managed elements such as computing/network/storage resources, and the four methods in REST can replace full operation of management systems. It also allows management systems to be easily decentralized because management information can modeled as a resource, which is identified by a URI.

*RESTful Cloud Management System (CMS)* [10] is composed of a GUI or external systems, a REST-based manager and a REST-based agent in the infrastructure element side. CMS fully utilizes fundamental Web technologies such as, HTTP and URIs, to perform infrastructure management through REST-based manager and agent.

3.2 Papaya

*Papaya* is an IaaS management platform proposed by Liutong et al. [7] is based on Eucalyptus, Libvirt management Library, SNMP (Simple Network Management Protocol), Virtual Machine Monitor (VMM) for managing and monitoring cloud. Papaya helps enterprises to manage and monitor physical resources, virtual machines, virtual machine images, virtual machines' status etc. in their private cloud. Papaya management platform can manage the VM's lifecycle, resize the VM's resources, manage other resources' lifecycle in the cloud like VM image, VM volume, key pairs etc. As a whole, Papaya can manage the IaaS based on Eucalyptus, collect both the real-time working-node's static and dynamical information and VM's information running on the node, which can provide a better reference for resource discovery, resource scheduling and load balancing and billing the use of resources and can be helpful to ensure the IaaS cloud's stability.

3.3 OpenNebula

*Virtual Infrastructure Management (VIM)*, the dynamic orchestration of VMs is a key component to provide users with the same features found in commercial public/private/hybrid clouds. The system VM provides a complete system platform that supports the execution of a complete operating system (OS). The VM life cycle has six phases: create, suspend, resume, save, migrate, and destroy. On a same physical node, a number of VMs having different OS can run simultaneously. A *Virtual machine Monitor (VMM)*, referred as *hypervisor* is used to control and manage the VMs on a single physical node. On top of VMM is the *Virtual Infrastructure Managers (VIMs)* are used to manage, deploy, and monitor VMs on a distributed pool of resources. *Cloud Infrastructure Managers (CIMs)* are web-based management solutions on top of IaaS providers [11]. VIM must provide a uniform and homogeneous view of virtualized resources regardless of underlying platform; manage a VMs full life cycle, support configuration resource allocation policies to meet the organization's changing resource needs; including peaks in which local resources are insufficient, and changing resources including addition or failure of resources.

*OpenNebula* proposed by Borja et al. [12] is an open source Virtual Infrastructure Manager that can be used by organizations to deploy and manage VMs, either individually or in groups that must be scheduled on local resources or external public clouds. It automates VM setup regardless of the underlying virtualization layer or external cloud. Haizea [12] is a resource lease manager that can act as a scheduling back end for OpenNebula, providing leasing capabilities not found in other cloud systems such as *advance reservations (ARs)* and *resource preemption* which are particularly relevant for private clouds. A key feature of OpenNebula's architecture is its highly modular design, which facilitates third party component in the cloud ecosystem, such as cloud toolkits, virtual image managers, service managers, and VM schedulers. By relying on a flexible, open, and loosely coupled architecture, OpenNebula is designed from the outset to be easy to integrate with other components, such as the Haizea lease manager. OpenNebula can support hybrid cloud model by using cloud drivers to interface with external clouds. When used together, OpenNebula and Haizea are the only Virtual Infrastructure Management solution providing leasing capabilities beyond immediate provisioning, including best-effort leases and advance reservation of capacity.

### 3.4 Other Infrastructure Management Cloud Projects

Other than OpenNebula, Papaya, CMS, the Cloud Infrastructure Management projects are; *Open Cloud Computing Interface (OCCI)* which uses RESTful protocol and API for all kinds of management tasks, the *Cloud Management Working Group (CMWG)* addresses management interoperability for clouds between service consumers and providers, *Cloud Data Management Interface (CDMI)* created by the Storage Network Industry Association (SNIA) enables interoperable cloud storage and data management.

## 4. Integrated Multi-Cloud Management Services

The cloud platforms must evolve from just infrastructure delivery to automated service to satisfy the full automation requirements demanded by service providers and it should be able to provide all the services through one gateway. Four goals to be accomplished for this are; appropriate service abstraction level, automatic scalability, smart scaling, and avoiding vendor lock-in. Cloud projects such as, *DeltaCloud,* and *RightScale* are some of the examples. DataCloud is an open source project, providing one common API for a wide range of service providers. RightScale provides management tools to manage the cloud infrastructure over multiple public cloud providers. Few of the models providing muti-cloud platform or multi-service platform are discussed in the following sections.

### 4.1 Multi Cloud Management Platform

Multi IaaS sites are started to appearing as cloud computing benefits spread. They require heavy workloads to migrate an application equipped with virtual server from one cloud site to another. To resolve the issues on multi Iaas, Tiancheng et al. proposes the *Multi Cloud Management Platform (MCMP)* [13] that locates between cloud users and cloud sites and provides unified cloud management portal for a server user and server administrator. It has mainly three features: service catalog federation, collaborative management, and application virtual server migration. Multi cloud environments become more useful using both MCMP and Virtual Private Cloud. Jamcracker and RightScale provide the capacity of connecting an enterprise with different cloud service providers.

### 4.2 Unified IaaS Proxy and Monsoon

*Unified IaaS Proxy* [14] is a common interface for the management of hybrid cloud proposed by Shixing et al. It is a generic abstraction model of IaaS services to facilitate multi-sourcing IaaS for increasing availability and security requirements. Various types of models in Unified IaaS Proxy are; *Resource type* such as compute and storage, *Reflection type* such as image and snapshot, *Credential* and *Firewall*. A number of services are proposed in the service model including; resource management service, virtual machine service, clone service, and security service. In a nutshell, Unified IaaS Proxy provides a common interface to manage IaaS environments across public clouds and private clouds, so the users only need to interface with IaaS proxy, but not with specific APIs from various IaaS providers. The IaaS

Proxy can be discovered and made available to other services running on web2Exchange and it also provides REST interface to the users and services situated outside of Web2Exchange. IaaS Proxy supports Amazon Web Service (AWS) EC2, GoGrid, Rackspace Cloud and private clouds such as HP CloudSystem Matrix.

*Monsoon* [14] portal sits on top of the Unified IaaS Proxy, and provides a multi-tenant infrastructure management portal for hybrid cloud. It also provides enterprise users a comprehensive management tool for their cloud infrastructure in hybrid cloud environment. In Monsoon portal, the requests to various IaaS providers across public and private clouds will be interpreted through the RESTful interface of IaaS Proxy, and the basic set of features provided by Monsoon to manage cloud infrastructure in hybrid cloud are; user management, access control, resource management, monitoring and reporting. Some of the advanced features provided by Monsoon are; corporate account, policy implementation, billing and charging. In short, Monsoon portal is developed to enable users to subscribe, monitor, and manage the full life-cycle of IaaS services from multiple providers across public and private clouds. Unified IaaS Proxy deployed in Monsoon is able to support AWS, GoGrid, Rackspace Cloud, HP BladeSystem Matrix and other IaaS providers like Openstack and Eucalyptus.

4.3 Claudia

Luis et al. proposed a new abstraction layer for cloud systems; *Claudia* that offers a more friendly interface to service providers by enabling the control of the service life cycle [15]. Claudia acts as a new abstraction layer on top of a Virtual Infrastructure Manager that enables the utilization of several clouds with heterogeneous interface from different cloud providers. Claudia uses OpenNebula Virtual Infrastructure Manager as the VIM and its features closes the gap between the service provider needs and are: single deployment operation, automatic, smart and diverse scalability, and addresses the issue of the vendor lock-in by providing seamless access to resources of different cloud providers for a single service.

4.4 Cloud@Home

IaaS Cloud Providers (CP) heavily relies on a virtualization system, as the leased computing resources are single or multiple Virtual Machines (VMs). *Cloud@Home (C@H)* proposed by Salvatore et al. aims at building an IaaS cloud provider using computing and storage resources acquired from volunteer contributions, also provides a system for managing the quality of the leased services [4]. C@H aims at delivering QoS on top of a physical infrastructure made up volunteered resources and it provides a set of tools enabling the composition of a new enhanced provider of resources namely C@H provider, which puts itself in acting as a resource aggregator and makes use of tools and internal services to evaluate and to predict the sustainability of the QoSrequested, as well as to identify the best combination of resources to be leased to the user.

C@H provider collects resources from several cloud providers (public/private) merging different technologies and heterogeneous resource management policies, and offers such resources to the users in a uniform way. The three actors in C@H provider are: Cloud Provider, C@H Admin, and C@H User. Functionalities supplied by C@H provider can be organized into three modules: resource abstraction, resource management, and SLA management modules. If a single C@H component is offered as a customized virtual machine hosted by any cloud provider, it is considered as PaaS, and the cloud system which offers basic building blocks composed to create customized C@H Provider supplying resources is an IaaS. As a whole, C@H offers desired level of QoS in the form of resource availability.

## 5. Conclusion

This paper gives a short review of the Cloud Computing, which is comparatively new and fascinating technology allows the cloud providers and users to be in a win-win situation, enjoying uninterrupted services with little infrastructure on the user end. Also, it discusses about the cloud infrastructures, infrastructure management models, the virtual infrastructure management etc. To provide the users all the services from one end, integrated multi-cloud management services and models have been introduced, which allows accessing the services from a single provider capable of integrating all the different providers and pool the resources on a single reservoir, with ease of use and higher QoS.

## References
[1] Peter Mell, Timothy Grance, "The NIST Definition of Cloud Computing (Draft)", NIST Special Publication 800-145, January 2011
[2] BorkoFurht, Armando Escalante, "Handbook of Cloud Computing", Springer Science+Business Media, 2010, e-ISBN 978-1-4419-6524-0
[3] BarrieSosinsky et al., "Cloud Computing Bible", Wiley India, 2011
[4] Salvatore Distefano, Antonio Puliafito, MassimilianoRak, Salvatore Venticinque, Umberto Villano, Antonio Cuomo, Giuseppe Di Modica, OrazioTomarchio, "QoS Management in Cloud@Home Infrastructures",

Proceedings of IEEE International Conference on Cyber-Enabled Distributed Computing and Knowledge Discovery, 2011, pp. 190 – 197

[5] Judith Hurwitz et al., "Cloud Computing for Dummies", Wiley, 2011

[6] Michael Maurer, Ivan Breskovic, Vincent C. Emeakaroha, and IvonaBrandic, "Revealing the MAPE Loop for the Autonomic Management of Cloud Infrastructures", Proceedings of IEEE MoCS'11, June 2011, pp. 147 – 169

[7] LiutongXu, Jie Yang, "A Management Platform for Eucalyptus-Based IaS", Proceedings of IEEE CCIS2011, pp. 193 – 197

[8] AhsanArefin, Guofei Jiang, "CloudInsight: Shedding Light on the Cloud", Proceedings of 30th IEEE International Symposium on Reliable Distributed Systems, 2011, pp. 219 – 228

[9] Roy T. Fielding, "Architectural Styles and the Design of Network-based Software Architecture", Doctoral dissertation, 2000

[10] Hyuck Han, Shingyu Kim, Hyungsoo Jung, Heon Y. Yeom, Changho Yoon, Jongwon Park, Yongwoo Lee, "A RESTful Approach to the Management of Cloud Infrastructure", Proceedings of IEEE International Conferenvc on Cloud Computing, IEEE Computer Society, 2009, pp. 139 – 142

[11] Nick Antonopoulos, Lee Gillam, "Cloud Computing – Principles, Systems and Applications", Springer-Verlag London Limited, 2010, e-ISBN 978-1-84996-241-4

[12] Borja Sotomayor, Ruben S. Montero, Ignacio M. Liorente, Ian Foster, "Virtual Infrastructure Management in Private and Hybrid Clouds", IEEE Internet Computing, Published by the IEEE Computer Society, September/October 2009

[13] Tiancheng Liu, Yasuharu Katsuno, Kewei Sun, Ying Li, Takayuki Kushida, Ying Chen, Mayumi Itakura, "Multi Cloud Management for Unified Cloud Services Across Cloud Sites", Proceedings of IEEE CCIS 2011, pp. 164 - 169

[14] Shixing Yan, Bu Sung Lee, Guopeng Zhao, Ding Ma, Peer Mohamed, "Infrastructure Management of Hybrid Cloud for Enterprise Users", 978-1-4577-1811-3/11, IEEE 2011

[15] Luis Rodero-Merino, Luis M. Vaquero, Victor Gil, Fermin Galan, Javier Fontan, Ruben S. Montero, Ignacio M. Llorente, "From Infrastructure Delovery to Service Management in Clouds", Future Feneration Computer Systems, 26 (2010), www.elsevier.com/locate/fgcs, pp. 1226 – 1240

**A. Anasuya Threse Innocent** achieved her Bachelor of Engineering in Electronics and Instrumentation in 2002 under Manonmanium Sundaranar University and her Master of Engineering in Computer Science and Engineering under Anna University in 2004 from Noorul Islam College of Engineering, Kumaracoil, Tamilnadu, India (currently Noorul Islam University). She also worked as a Lecturer there for one year. Then she worked as an Assistant Professor in the School of Information Science and Engineering, VIT University, Vellore, Tamilnadu, India for about 6 years. Currently she is working as an Assistant Professor and Head of the Departments of Computer Science and Engineering and Information Science and Engineering in SCT Institute of Technology, Bangalore under Visvesvaraya Technological University, Karnataka, India. She has about 11 International and National Journal and Conference papers. Also she got best paper award for one of her National Conference papers. Her research areas of interest include; Cloud Computing, Cryptography, Data Compression etc. She is also a life member of CSI.